\documentclass[12pt,a4paper]{article}
\usepackage[utf8]{inputenc}
\usepackage{amsmath,bbm,longtable,paralist}
\usepackage{caption}
\usepackage[top=1.4in, bottom=1.4in, left=1in, right=1in]{geometry}

\DeclareCaptionStyle{italic}[justification=centering]{labelfont={bf},textfont={it},labelsep=colon}
\usepackage{graphicx,psfrag,epsf}
\newcommand{\coloneqq}{\mathrel{\mathop:}=}
\usepackage{enumerate}
\usepackage{natbib}
\usepackage{tikz}
\usetikzlibrary{shapes.geometric, arrows.meta, positioning}

\usepackage{url}
\usepackage{bbm, dsfont}
\usepackage{comment}
\usepackage{geometry}
\usepackage{float}
\usepackage[dvipsnames, table]{xcolor}
\usepackage{booktabs, subfig, bm, paralist,mathpazo,tikz,todonotes,longtable,microtype,dsfont,rotating} 
\usepackage[pdftex,colorlinks=true]{hyperref}
\definecolor{darkblue}{rgb}{0,0,.6}
\hypersetup{citecolor=darkblue,linkcolor=darkblue,urlcolor=darkblue}

\newcommand{\blind}{0}
\usepackage{algorithm}
\usepackage{algorithmic}

\graphicspath{{plots/}}

\newsavebox\CBox

\definecolor{a0}{rgb}{0.0, 0.5, 0.0}
\definecolor{bistre}{rgb}{0.24, 0.17, 0.12}
\definecolor{amethyst}{rgb}{0.6, 0.4, 0.8}
\definecolor{blue-violet}{rgb}{0.54, 0.17, 0.89}
\definecolor{Rcolor}{RGB}{150,160,190}
\definecolor{blush}{rgb}{0.87, 0.36, 0.51}
\definecolor{brightturquoise}{rgb}{0.03, 0.91, 0.87}
\definecolor{burntorange}{rgb}{0.8, 0.33, 0.0}

\usepackage{orcidlink}
\date{}
\AtBeginDocument{}

\begin{document}

\def\spacingset#1{\renewcommand{\baselinestretch}{#1}\small\normalsize} \spacingset{1}

\if0\blind
{
  \title{\bf A Beta-GAM Hidden Markov Model for Proportion Time Series}
  \author{\normalsize Andrea Nigri\footnote{Corresponding author, email: andrea.nigri@unifg.it} \orcidlink{0000-0002-2707-3678}\\  
  \normalsize Department of Social Sciences \\
  \normalsize University of Foggia, Italy \\
  \\
  \normalsize Han Lin Shang \orcidlink{0000-0003-1769-6430}\\
   \normalsize Department of Actuarial Studies and Business Analytics \\
   \normalsize Macquarie University, Australia \\
   \\
\normalsize Marco Bonetti \orcidlink{0000-0003-2304-4180} \\
\normalsize Department of Social and Political Sciences  \\
\normalsize Bocconi University, Italy
}
  \maketitle
} \fi

\if1\blind
{
\title{\bf A Hidden Markov Model with Beta-GAM Emissions}
  \maketitle
} \fi

\bigskip

\begin{abstract}
We propose a hidden Markov model for univariate proportion time series taking values in $(0,1)$, where regime switching captures latent structural changes and the emission distribution belongs to the Beta family. In each latent state, the Beta mean is linked to covariates through a generalized additive model (GAM) with spline-based smooth functions, while the Beta precision is state-specific, enabling flexible modeling of both nonlinear covariate effects and regime-dependent variability. Estimation is carried out via a penalized expectation--maximization algorithm, combining smoothing with numerical maximization of the penalized emission likelihood. To select the number of latent states and the smoothing penalty, we implement a grid search guided by standard information criteria (Akaike Information Criterion/Bayesian Information Criterion/Integrated Completed Likelihood) with a diagnostic filter that removes degenerate solutions characterized by explosive precision estimates. Uncertainty is quantified through a parametric bootstrap procedure for transition probabilities and state-dependent parameters. Simulation results demonstrate accurate recovery of transition dynamics, state precisions, and latent-state decoding. A motivating application to Russian age-specific mortality data (1960--2014, ages 0--40) illustrates how the proposed model summarizes smooth age patterns in female-to-total mortality ratios while identifying two persistent latent regimes that admit a substantive demographic interpretation in light of the country's well-documented mortality shocks that occurred over the second half of the twentieth century. 

\noindent \textbf{Keywords:} Beta regression; Compositional time series; EM algorithm; Generalized additive model; Regime switching model
\end{abstract}

\newpage
\spacingset{1.6}

\section{Introduction}\label{sec:1}

The presence of proportion data is common in economics, finance, and social statistics, including the income distribution \citep{KU01}, financial return distribution \citep{KMP+19}, and the distributions of times when bids are submitted in an online auction \citep{JCP+08}. In demography, one can model a redistribution of normalized life-table death counts over time \citep[see, e.g.,][]{PLC19, BKC20}. 

More recently, Beta-distributed emissions have been studied in lockstep with a Hidden Markov model (HMM) and regime-switching additive predictors. On the one hand, \cite{CES22} proposed a Bayesian Beta-HMM for univariate proportion time series, using conjugate priors and Markov Chain Monte Carlo methods for posterior and predictive inference. Their contribution provides an important starting point for Beta-distributed HMMs, but the emission parameters are constant within each state, and no covariate-dependent structure is introduced. On the other hand, \cite{LKGM17} developed Markov-switching generalized additive models (MS-GAMs), showing how state-dependent smooth covariate effects can be estimated by combining HMM with penalized B-splines and maximum penalized likelihood. However, that framework is not specifically designed for Beta-distributed proportion data. Our paper is positioned at the intersection of these two contributions: we combine the flexibility of Markov-switching GAMs with a Beta emission model tailored to bounded responses in $(0,1)$. Specifically, we introduce a matrix-structured Hidden Markov Model (HMM) to model time series of univariate percentage data. The model features Beta-distributed emissions, where the distribution parameters are governed by a generalized additive model (GAM), thus allowing for smooth dependence on covariates. Each hidden state in the Markov chain determines a distinct Beta distribution, with both the mean (linked via a logistic transformation) and the precision (a scalar) being state-specific. This formulation is particularly suitable for modeling rates, probabilities, or proportions \citep[see, e.g.,][]{FC04, DW19}.

Compared to \cite{CES22}, our model allows the conditional mean of the Beta distribution to vary smoothly with observed covariates within each latent state, while retaining a state-specific precision parameter. Relative to \cite{LKGM17}, we specialize the regime-switching additive framework to proportion time series by adopting a Beta mean -- precision parametrization, which is natural for rates, shares, and ratios. Methodologically, the contribution is four-fold: 
\begin{asparaenum}
\item[(i)] We embed a generalized additive model (GAM) with B-spline basis expansions within the Beta emission distribution, allowing the mean to vary nonlinearly across covariates and across latent regimes. 
\item[(ii)] We adopt a frequentist penalized expectation-maximization (EM) estimation strategy rather than a fully Bayesian approach, which is computationally attractive for large samples and avoids the specification of prior distributions for high-dimensional spline coefficient vectors. 
\item[(iii)] We introduce a diagnostic criterion to detect degenerate solutions when the number of latent states is overspecified, and incorporate this criterion into a grid-search-based model selection procedure. 
\item[(iv)] Beyond the algorithmic proposal itself, we provide explicit analytical derivations for the complete-data likelihood and for the derivatives of the penalized $\log$-likelihood underlying the estimation routine. 
\end{asparaenum}
For this combined Beta-GAM HMM setting, these derivations supply a formal mathematical foundation for the proposed estimation procedure and clarify how the EM updates are constructed.

In our proposal, we study proportions that are clearly bounded within the unit interval. Let $\{(y_1, y_2,\dots, y_T), y_t\in (0,1)\}$ be a univariate time series of proportions. For each proportion, we also observe a set of covariates $\bm{x}_t$. To model possibly nonlinear relationships, we let, conditionally on the latent state $z_t = k$, the response follows a Beta distribution with state-specific precision and a mean linked to covariates through a logistic function,
\[
y_t \mid z_t = k \sim \mathrm{Beta}(\mu_{kt}\phi_k,\,(1-\mu_{kt})\phi_k), 
\qquad \mathrm{logit}(\mu_{kt}) = f_k(\bm{x}_t),
\]
where each $f_k(\cdot)$ is a smooth function estimated via B-spline 
basis expansions.

Due to its flexibility, we resort to a GAM of \cite{Wood17} and its estimation variants \citep[see, e.g.,][]{HYS+19}. The unknown function $f(\cdot)$ can be estimated by using $B$-spline functions and their associated coefficients. These coefficients are assumed to be realized with several latent states. In this way, the proposed model can be viewed as a Beta-distributed, covariate-dependent extension of the Beta-HMM in \cite{CES22}, and at the same time as a proportion-specific specialization of the MS-GAM framework in \cite{LKGM17}. We estimate the optimal number of latent states and the smoothing penalty through information criteria combined with diagnostic filtering, and use penalized EM algorithm for estimating the state-dependent mean and precision parameters.

The outline of the paper is structured as follows. In Section~\ref{sec:2}, we present a HMM framework within the Beta family. The joint distribution of the latent states and the proportions is discussed. In Section~\ref{sec:3}, we present the $B$-spline approximation for the estimation of the unknown function $f(\cdot)$, where a roughness penalty is placed on the coefficients of the $B$-splines. In Section~\ref{sec:4}, we develop an EM algorithm and a bootstrap to quantifying the uncertainty in parameter estimation. Section~\ref{Sec:5} introduces diagnostics-based model selection and the associated uncertainty quantification strategy. For completeness, the analytical derivatives that support the penalized estimation routine are collected in Appendix~\ref{sec:Appendix_A}. Leveraging a series of simulation studies in Section~\ref{sec:6}, we investigate the finite-sample properties of the parameter estimation method. Section~\ref{sec:7} presents the empirical study investigating a time series of ratios between female and total mortality. Conclusions are given in Section~\ref{sec:8}, along with some ideas on how the methodology can be further extended.

\section{Hidden Markov Model Framework}\label{sec:2}

A HMM provides a flexible probabilistic structure for modeling time series of observed data $(y_1, y_2,\dots, y_T)$, assumed to arise from an underlying unobserved sequence of latent states $z_t \in \{1, 2, \ldots, K\}$. Each observation may depend on a vector of covariates $\bm{x}_t \in \mathbb{R}^p$. The sequence of hidden states $(z_1, z_2, \ldots, z_T)$ is assumed to evolve according to a first-order Markov chain, with an initial distribution $\bm{\pi} = (\pi_1, \pi_2, \ldots, \pi_K)^{\top}$, where $\pi_k = P(z_1 = k)$, satisfying 
\begin{equation*}
\sum_{k=1}^K \pi_{k} = 1, \quad \pi_k \geq 0. 
\end{equation*}
The evolution of the latent state sequence is governed by the $K \times K$ transition probabilities matrix $\boldsymbol{A} = [A_{ij}]$, where the $(i,j)$\textsuperscript{th} entry represents the probability of transitioning from state $i$ at time $t-1$ to state $j$ at time $t$, for all times $t\in \{2,3,\dots,T\}$.
\begin{equation*}
A_{ij} = P(z_t = j \mid z_{t-1} = i), \qquad i, j \in \{1, 2, \ldots, K\}.
\end{equation*}
By construction, $\boldsymbol{A}$ is a row-stochastic matrix, which means that each row sums to one, that is, for $i\in \{1,2,\dots,K\}$ one has
\begin{equation*}
\sum_{j=1}^{K} A_{ij} = 1, \qquad A_{ij} \ge 0 \quad \forall i, j.
\end{equation*}
The matrix $\boldsymbol{A}$ encodes the Markov dynamics of the latent process: starting from an initial state $z_1$ drawn according to $\pi$, the probability of observing a particular sequence of states $(z_1, z_2, \ldots, z_T)$ is given by
\begin{equation*}
P(z_1, \ldots, z_T) = {\pi_{z_1} } \prod_{t=2}^T A_{z_{t-1}, z_t}.
\end{equation*}
Each element $A_{ij}$ characterizes the switching behavior between regimes, allowing persistence within states (when $i = j$) and transitions between different states (when $i \neq j$). 

\subsection{Beta Emission Model}\label{sec:2.1}

Let $y_t \in (0, 1) \subset \mathbb{R}$ be a percentage observed at time $t$, and $\bf{x}_t \in \mathbb{R}^{p}$ the covariate vector. Let $z_t \in \{1, \ldots, K\}$ denote the latent state at time $t$. Conditionally on the latent state $z_t = k$, we assume
\begin{equation*}
y^{(k)}_t \coloneqq y_t|z_t=k  \sim \text{Beta}(a_{kt}, b_{kt}),
\end{equation*}
with parameters defined as
\begin{equation*}
a_{kt} = \mu_{kt} \phi_k, \qquad b_{kt} = (1 - \mu_{kt}) \phi_k,
\end{equation*}
where $\mu_{kt} \in (0, 1)$ is the mean and $\phi_k > 0$ is a state-specific precision parameter. The Beta density is given by
\begin{equation*}
f(y_t \mid a_{kt}, b_{kt}) = \frac{y_t^{a_{kt} - 1}(1 - y_t)^{b_{kt} - 1}}{B(a_{kt}, b_{kt})}, \quad y_t \in (0,1),
\end{equation*}
where $B(a, b) = \dfrac{\Gamma(a) \Gamma(b)}{\Gamma(a + b)}$ is the Beta function. The mean $\mu_{kt}$ is related to an additive predictor $\eta_{kt}$ through the logit function,
\begin{equation*}
\mu_{kt} = \frac{1}{1 + \exp^{-\eta_{kt}}}, \quad \text{with} \quad \eta_{kt} = \sum_{j=1}^p f_{kj}(x_{tj}),
\end{equation*}
where each $f_{kj} : \mathbb{R} \to \mathbb{R}$ is a smooth function represented via basis expansions.

\subsection{Joint Distribution}\label{sec:2.2}

Let $\bm{z} = (z_1, z_2, \ldots, z_T)^{\top}$ denote the sequence of latent states, and $\bm{y} = (y_1, y_2, \ldots, y_T)^{\top}$ the observed data. The joint likelihood of complete data with parameters $\Theta = (\bm{\pi}, \boldsymbol{A}, \{\beta_k\}, \{\phi_k\})$ is given by 
\begin{equation*}
P(\boldsymbol{y}, \boldsymbol{z} \mid \Theta) =P(\boldsymbol{y}, \boldsymbol{z} \mid \bm{\pi}, \boldsymbol{A}, \{\beta_k\}, \{\phi_k\}) = \pi_{z_1} \prod_{t=2}^T A_{z_{t-1}, z_t} \prod_{t=1}^T p(y_t \mid z_t, x_t).
\end{equation*}
The marginal likelihood is obtained by summing over all possible latent state sequences:
\begin{equation*}
P(y \mid \Theta) = \sum_{z_{1:T} \in \{1, \ldots, K\}^T} P(\boldsymbol{y}, \boldsymbol{z} \mid \Theta),
\end{equation*}
which is computed using the forward algorithm. Analytical derivatives of the penalized $\log$-likelihood are reported in Appendix~\ref{sec:Appendix_A}.

\section{Functional Representation via Additive Predictors}\label{sec:3}

With a slight change in notation, we now let $\bm{x}_{t\bullet} = (x_{t1}, \ldots, x_{tp})^\top \in \mathbb{R}^{p}$ denote the covariate vector \footnote{All numerical results presented here consider the case of a single covariate ($p=1$), which suffices for both the simulation design and the mortality application in Section~\ref{sec:7}. The extension to multiple covariates requires only the concatenation of per-covariate basis matrices into $\mathcal{B}$ as described above, and involves no changes to the estimation algorithm.} at time $t$, and let $\bm{x}_{\bullet j} = (x_{1j},\dots, x_{Tj})^{\top}~\in R^{T}$ denote the vector of covariate values for variable $j$. For each state \( k \in \{1, \ldots, K\} \), the additive predictor is defined as:
\begin{equation}
\eta_{kt} = \sum_{j=1}^p f_{kj}(x_{tj}), \label{eq:sec_3}
\end{equation}
where each \( f_{kj} \subset L^2(\mathbb{R}) \) is a smooth function approximated using a spline basis expansion:
\begin{equation*}
f_{kj}(x_{tj}) = \sum_{m=1}^M \beta_{kjm} B_m(x_{tj}),
\end{equation*}
with $\{B_m(x_{tj})\}_{m=1}^{M}$ a set of $M$ basis functions (e.g. $B$-splines), where $M$ is the basis dimension, and \(\beta_{kjm} \in \mathbb{R} \) the corresponding coefficients. Following the P-spline framework of \cite{EM96}, we use cubic B-splines with a moderately large number of basis functions and control smoothness through a discrete difference penalty on the coefficients rather than through the number of knots.

Define the design matrix with a specific basis spline for covariate $j$ as:
\begin{equation*}
\bm{B(\bm{x}_{\bullet j})} = 
\begin{bmatrix}
B_1(x_{1j}) & \cdots & B_M(x_{1j}) \\
\vdots & \ddots & \vdots \\
B_1(x_{Tj}) & \cdots & B_M(x_{Tj})
\end{bmatrix} \in \mathbb{R}^{T \times M},
\end{equation*}
and concatenate all covariate basis matrices and coefficients:
\begin{equation*}
\mathcal{B} = [B_1 \mid B_2 \mid \cdots \mid B_p] \in \mathbb{R}^{T \times pM}, \qquad
\beta_k = 
\begin{bmatrix}
\beta_{k1}^\top & \cdots & \beta_{kp}^\top
\end{bmatrix}^\top \in \mathbb{R}^{pM},
\end{equation*}
with $\beta_{kj} = (\beta_{kj1}, \ldots, \beta_{kjM})^\top \in \mathbb{R}^{M}$.

The predictor vector over all times is then:
\begin{equation*}
\bm{\eta}_k = \bm{\mathcal{B}} \bm{\beta}_k \in \mathbb{R}^{T}, \quad \text{and} \quad \mu_{kt} = \frac{1}{1 + \exp^{-\eta_{kt}}},
\end{equation*}
with $\eta_{kt}$ as in~\eqref{eq:sec_3}.

\section{Details of the Expectation-Maximization Algorithm}\label{sec:4}

We estimate the set of parameters $\Theta = (\bm{\pi}, \bm{A}, \{\bm{\beta}_k\}_{k=1}^{K}, \{\phi_k\}_{k=1}^{K})$ via the EM algorithm, which iteratively maximizes the expected $\log$-likelihood of the complete data. To encourage smoothness in the estimated functions, a quadratic penalty is applied to the spline coefficients, leading to a penalized maximum likelihood formulation.

\subsection{E-step: Forward-Backward Recursion}\label{sec:4.1}

Define the forward variables: 
\begin{align*}
L_1(k) &= \pi_k \, P(y_1 \mid z_1 = k), \\
L_t(j) &= P(y_t \mid z_t = j) \sum_{i=1}^K L_{t-1}(i) A_{ij}, \quad t = 2, \ldots, T.
\end{align*}

Define the backward variables:
\begin{align*}
R_T(k) &= 1, \\
R_t(i) &= \sum_{j=1}^K A_{ij} \, P(y_{t+1} \mid z_{t+1} = j) R_{t+1}(j), \quad t = T-1, \ldots, 1.
\end{align*}

The smoothed posterior probabilities are:
\begin{equation*}
\gamma_{tk} = \frac{L_t(k) R_t(k)}{\sum_{l=1}^K L_t(l) R_t(l)}.
\end{equation*}
In practice, to avoid numerical underflow for long time series, all forward-backward computations are carried out on the $\log$ scale using the $\log$-sum-exp identity $\log\!\sum_i e^{a_i} = a_{\max} + \log\!\sum_i e^{a_i - a_{\max}}$, following the standard HMM scaling approach described in  \cite{ZML16}.

The joint posterior of successive states is:
\begin{equation*}
\xi_{tij} = \frac{L_t(i) A_{ij} P(y_{t+1} \mid z_{t+1} = j) R_{t+1}(j)}{
\sum_{i'=1}^K \sum_{j'=1}^K L_t(i') A_{i'j'} P(y_{t+1} \mid z_{t+1} = j') R_{t+1}(j')}.
\end{equation*}

\subsection{M-step: Maximization of the Q-function}\label{sec:4.2}

The expected complete-data $\log$-likelihood is maximized with respect to each component of $\Theta$.

\paragraph{Emission $\log$-likelihood.}

The $\log$-likelihood contribution for a single observation, conditional on \( z_t = k \), is:
\begin{align*}
\log p(y_t \mid z_t = k) =& \log \Gamma(\phi_k) - \log \Gamma(\mu_{kt} \phi_k) - \log \Gamma((1 - \mu_{kt}) \phi_k) + (\mu_{kt} \phi_k - 1) \log y_t + \\
& ((1 - \mu_{kt}) \phi_k - 1) \log(1 - y_t).
\end{align*}

\paragraph{Update of $\beta_k$.}

To prevent overfitting, we introduce the quadratic penalty term. The penalized objective function is therefore
\begin{equation*}
\widetilde{Q}(\beta_k) = \sum_{t=1}^{T} \gamma_{tk} \log P(y_t \mid \beta_k, \phi_k) - \lambda \beta_k^\top \mathbf{P} \beta_k,
\end{equation*}
where $\mathbf{P} = \mathbf{D}_d^\top \mathbf{D}_d$ is the penalty matrix constructed from the $d$th-order difference matrix $\mathbf{D}_d$ applied to adjacent B-spline coefficients \citep{EM96}. With $d=2$, this penalizes the integrated squared second derivative of the fitted curve, encouraging smoothness while preserving the ability to capture nonlinear patterns. As $\lambda\to\infty$, the penalty forces the coefficients to lie on a straight line, so that the common parametric linear predictor is recovered as the limiting case \citep{LKGM17}.

Let $\nabla \mu_{kt} = \mu_{kt}(1 - \mu_{kt})$, and let $b(x_t)$ denote the basis vector in time $t$. The gradient of $\widetilde{Q}$ is:
\begin{equation*}
\nabla_{\beta_k} \widetilde{Q} = \sum_{t=1}^T \gamma_{tk} \, \phi_k 
\left( \log \frac{y_t}{1 - y_t} - \psi(\mu_{kt} \phi_k) + \psi((1 - \mu_{kt}) \phi_k) \right) 
\nabla \mu_{kt} \, b(x_t) - 2 \lambda \mathbf{P} \beta_k.
\end{equation*}

The updates of $\beta_k$ and $\phi_k$ are performed jointly by numerically minimizing the negative penalized $Q$-function with respect to $(\beta_k, \log\phi_k)$ via the L-BFGS-B algorithm \citep{NW06}, treating each state independently. Since these updates 
do not guarantee a closed-form maximum of the complete-data $\log$-likelihood, the procedure constitutes a generalized EM algorithm \citep{DLR77}, in which monotone increase of the penalized observed-data log-likelihood is preserved at each iteration. In the remainder of the paper we refer to this procedure as the \emph{penalized EM} algorithm, keeping in mind that, technically, it is a generalized EM because the M-step maximizes the penalized $Q$-function only numerically rather than in closed form.

The precision parameter $\phi_k$ is re-parametrized as $\log\phi_k$ and optimized jointly with $\beta_k$. The analytical gradient with respect to $\phi_k$ is reported in Appendix~\ref{sec:Appendix_A}.

\section{Diagnostics-Based Selection Criterion for the Number of Latent States}\label{Sec:5}

Model selection in Hidden Markov Models (HMMs) requires the specification of both the number of latent states $K$ and any regularization parameters involved in emission modeling.

Model selection is traditionally performed via information criteria such as the Bayesian Information Criterion (BIC) or Integrated Completed Likelihood (ICL) criterion. However, when using flexible emission distribution, such as the Beta distribution, these criteria may favor overfitted models, introducing redundant latent states that may inflate the complexity of the model.

Indeed in our setting, overestimation of the number of latent states \( K \) may lead to degenerate solutions with some states receiving negligible probability, and their estimated Beta precision parameters \( \{\phi_k\}_{k=1}^K \) diverging toward the upper boundary of the parameter space. To mitigate this problem, we introduce a criterion that aims at excluding such degenerate solutions from consideration.

\paragraph{Diagnostic Criterion}

Let $\{\phi_k\}_{k=1}^{K}$ be the estimated precision parameters, and let $\phi_{(1)} \leq \phi_{(2)} \leq \dots \leq \phi_{(K)}$ denote the ordered values. 

Let $\phi_{\max}$ be the maximum value allowed during optimization. This bound is an implementation hyperparameter, indeed, it must be set large enough to comfortably exceed the plausible magnitude of the true precisions for the data at hand, so that saturation indicates genuine degeneracy rather than an artificial restriction of the parameter space. In the simulation study of Section~\ref{sec:6} we use $\phi_{\max} = 500$; in the mortality application of Section~\ref{sec:7}, where the observed proportions are highly concentrated, we use $\phi_{\max} = 2{,}000$ (see Section~\ref{sec:6}).

Define the differences \( \Delta_i = \phi_{(i+1)} - \phi_{(i)} \) for \( i = 1, \dots, K-1 \).

We define two diagnostic tests:
\begin{equation*}
N_{\text{sat}} = \sum_{k=1}^K \mathbb{1}\left( \phi_k \geq \phi_{\max} - \varepsilon\right)
\end{equation*}
The model is flagged if \( N_{\text{sat}} \geq s_{\text{thresh}} \), where typically \( \varepsilon = 10^{-6} \) and \( s_{\text{thresh}} = 2\).

To detect abrupt explosions in the upper tail of the ordered precisions, let
\[
m = \min(3, K-1),
\qquad
\Delta_{\text{tail}} = \sum_{i=K-m}^{K-1} \Delta_i.
\]
Let $\Delta_{\text{abs}}$ denote a per-jump threshold and $\Delta_{\text{sum}}$ a cumulative threshold.\footnote{Throughout the paper we used $\varepsilon = 10^{-6}$ and $s_{\text{thresh}} = 2$. The scale-dependent thresholds $(\phi_{\max}, \Delta_{\text{abs}}, \Delta_{\text{sum}})$ are set per application: $(500, 50, 100)$ in the simulation study of Section~\ref{sec:6} and $(2{,}000, 500, 1{,}000)$ in the mortality application of Section~\ref{sec:7}.}
The model is flagged if either
\[
\text{(i) } \exists i \in \{K-m,\ldots,K-1\}\ \text{s.t.}\ \Delta_i > \Delta_{\text{abs}}
\qquad \text{or} \qquad
\text{(ii) } \Delta_{\text{tail}} > \Delta_{\text{sum}}.
\]

Lastly, define the explosion indicator
\begin{equation*}
\mathcal{E}(K) =
\begin{cases}
1, & \text{if } N_{\text{sat}} \geq s_{\text{thresh}}
\ \text{or}\ 
\max_{i \in \{K-m,\ldots,K-1\}} \Delta_i > \Delta_{\text{abs}}
\ \text{or}\ 
\Delta_{\text{tail}} > \Delta_{\text{sum}}, \\
0, & \text{otherwise}.
\end{cases}
\end{equation*}
and retain only models with $\mathcal{E}(K) = 0$.

In addition to precision-based diagnostic, we impose a minimum state occupancy requirement: a model is discarded if any state receives less than a threshold share (e.g.\ 5\%) of the total posterior mass $\sum_t \gamma_{tk}$. This guards against solutions in which a nominally non-degenerate state captures only a handful of observations, leading to poorly identified emission parameters.

This diagnostic criterion complements standard model selection by removing structurally degenerate fits, often arising in high-dimensional or low-sample scenarios. 

These thresholds are conservative and effective across a range of settings. Applying this diagnostic substantially reduced the instability in model fitting, promoted interpretability, and favored more parsimonious solutions.

We now turn to model selection.

\subsection{Model Selection via Penalized Grid Search over Values of \texorpdfstring{$(K, \lambda)$}{(K, lambda)}}

To jointly select the optimal number of states $K$ and the smoothing parameter $\lambda$, we employ a grid search strategy over a finite set of candidate values:
\begin{equation*}
(K, \lambda) \in \mathcal{G}:=\mathcal{K} \times \Lambda,
\end{equation*}
where $\mathcal{K}=\left\{2,3, \ldots, K_{\max }\right\}$ is a set of possible numbers of states, and $\Lambda=\left\{\lambda_1, \ldots, \lambda_L\right\}$ is a set of smoothing penalty values.

For each combination $(K, \lambda)$, the model is fitted via penalized EM, with multiple random starts to mitigate local optima. The best solution (in terms of $\log$-likelihood) is retained for further evaluation.

\paragraph{Information Criteria:} Given the set of parameters $\widehat{\theta}$, to compare models across the grid we compute the following model selection criteria: replacing the raw parameter count $d_{\mathrm{par}}$ with the effective degrees of freedom $\nu$ (see below):
\begin{itemize}    
\item AIC$_\nu$:
\begin{equation*}
\mathrm{AIC}_\nu=-2 \ell(\widehat{\theta})+2\nu,
\end{equation*}
where $\nu$ is the effective number of parameters, accounting for the shrinkage induced by the penalty \citep{Akaike74}.
\item BIC$_\nu$:
\begin{equation*}
\mathrm{BIC}_\nu=-2 \ell(\widehat{\theta})+\nu \log T,
\end{equation*}
where $T$ is the number of observations (time points) \citep{Schwarz78}.
\item ICL$_\nu$:
\begin{equation*}
\mathrm{ICL}_\nu = \mathrm{BIC}_\nu - 2\sum_{t=1}^T \sum_{k=1}^K 
\gamma_{tk} \log \gamma_{tk},
\end{equation*}
where $\gamma_{t k}$ denotes the posterior probability of being in state $k$ at time $t$, as computed in the E-step. The ICL penalizes the model's entropy, favoring solutions with clearer state assignments \citep{BCG00}.
\end{itemize}

These criteria balance model fit and complexity, with BIC and ICL being more conservative and tending to favor more parsimonious models, while AIC may select more complex specifications.

\color{black}

In penalized models the nominal parameter count $d_{\mathrm{par}}$ overstates the effective model complexity, because the penalty shrinks the spline coefficients. Following 
\cite{Gray92} and \cite{LKGM17}, we replace the raw count with the effective degrees of freedom (EDF). For each state $k$, we approximate the EDF as
\[
\nu_k
=
\mathrm{tr}\!\left[
\left(\mathbf{B}^\top \mathbf{W}_k \mathbf{B} + \lambda \mathbf{P}\right)^{-1}
\mathbf{B}^\top \mathbf{W}_k \mathbf{B}
\right],
\]
where $\mathbf{W}_k = \mathrm{diag}(\gamma_{tk}\,\dot{\mu}_{kt})$ is a diagonal weight matrix derived from a local quadratic approximation of the penalized $Q$-function. This corresponds to the hat-matrix trace of a penalized weighted least-squares step evaluated at the current parameter estimates. Since the actual M-step is performed via direct numerical optimization of the full penalized $Q$-function (Section~\ref{sec:4.2}), rather than through iteratively reweighted 
least squares, this EDF is an approximation; it nonetheless provides a convenient and widely used scalar summary of effective model complexity 
that accounts for the shrinkage induced by the penalty \citep{Gray92, Wood17}. The total EDF 
\[
\nu = \sum_{k=1}^K \nu_k + K + K(K-1), 
\]
where the last two terms count the precision and transition parameters, then enters the corrected criteria $\mathrm{AIC}_\nu$ and $\mathrm{BIC}_\nu$.

\subsection{Two-stage selection}\label{sec:5.2}

We adopt a two-stage procedure for selecting $(K,\lambda)$: 
\begin{inparaenum}
\item[(i)] In the first stage, for each candidate $K$, the smoothing parameter $\lambda$ is chosen by minimizing $\mathrm{AIC}_\nu$ over $\Lambda$, since $\mathrm{AIC}_\nu$ is better suited for the continuous tuning of smoothness \citep{LKGM17}. 
\item[(ii)] In the second stage, the number of states $K$ is selected using $\mathrm{BIC}_\nu$ across the surviving $(K,\lambda^*(K))$ pairs, with preference for the richer state structure when competing models fall within two BIC units. 
\end{inparaenum}
This separation mirrors the common practice in GAMs of selecting the smoothing parameter by a less conservative criterion while using a stricter criterion for structural model choices.

Only models satisfying the diagnostic test $\mathcal{E}(K)=0$ are retained, and information criteria are compared only across the non-degenerate subset of $\mathcal{G}$, which we define as $\mathcal{G}_{\text {valid }}:=\{(K, \lambda) \in \mathcal{G} \mid \mathcal{E}(K)=0\}$. 

We adopt a two-stage selection procedure. In the first stage, for each candidate number of states $K$, the smoothing parameter is selected as
\[
\lambda^*(K) = \arg\min_{\lambda:\,(K,\lambda)\in\mathcal{G}_{\text{valid}}} \mathrm{AIC}_\nu(K,\lambda),
\]
where $\mathrm{AIC}_\nu$ uses the effective degrees of freedom $\nu$ in place of the raw parameter count. 

In the second stage, let
\[
\mathrm{BIC}_{\nu,\min}
=
\min_{K\in\mathcal{K}}
\mathrm{BIC}_\nu\bigl(K,\lambda^*(K)\bigr).
\]
We then define the admissible set
\[
\mathcal{K}_{\mathrm{tol}}
=
\left\{
K\in\mathcal{K}:
\mathrm{BIC}_\nu\bigl(K,\lambda^*(K)\bigr)
\le
\mathrm{BIC}_{\nu,\min}+2
\right\},
\]
and select
\[
K^*=\max \mathcal{K}_{\mathrm{tol}}.
\]

Thus, the smoothing parameter is chosen by $\mathrm{AIC}_\nu$ within each $K$, while the structural choice of $K$ is based on $\mathrm{BIC}_\nu$, with preference for the richer state structure when competing models are within two BIC units. In Appendix~\ref{sec:Appendix_B} we provide a schematic view of the selection procedure.

\color{black}

\subsection{Bootstrap Confidence Intervals for Model Parameters}\label{sec:5.3}

To quantify the uncertainty associated with the estimated parameters of the HMM Beta-GAM model, we construct confidence intervals using a parametric bootstrap procedure \citep[see, e.g.,][]{ZML16}. This approach captures both the variability due to the finite sample size and the inherent complexity of the model structure.

Let $\widehat{\Theta} = (\widehat{\bm{\pi}}, \widehat{\mathbf{A}}, \{\widehat{\beta}_k\}_{k=1}^K, \{\widehat{\phi}_k\}_{k=1}^K)$ be the parameter estimates obtained from the EM algorithm. The bootstrap procedure proceeds as follows. For each of the $B$ replications, a new dataset is generated by simulating latent state sequences $\{z_t^*\}_{t=1}^T$ from the fitted Markov chain, starting from $z_1^* \sim \widehat{\bm{\pi}}$ and evolving according to the transition matrix $\widehat{\mathbf{A}}$. Depending on the simulated state sequence and the covariate values, synthetic responses $\{y_t^*\}_{t=1}^T$ are then generated from the Beta emission model, using the estimated state-dependent mean and precision parameters. Specifically, for each $t$ and state $k$, the additive predictor $\eta_{kt}^*$ is computed using the estimated spline coefficients and the corresponding mean $\mu_{kt}^* = (1 + \exp(-\eta_{kt}^*))^{-1}$. The synthetic observation is then drawn as $y_t^* \sim \mathrm{Beta}(\mu_{z_t^*,t}^* \widehat{\phi}_{z_t^*},\ (1-\mu_{z_t^*,t}^*) \widehat{\phi}_{z_t^*})$.

For each bootstrap dataset, the full estimation procedure is repeated to obtain new parameter estimates $\widehat{\Theta}^*$. Collecting the resulting bootstrap samples over all $B$ replicates yields empirical distributions for each model parameter. The percentile-based confidence intervals are then constructed by extracting the empirical quantiles from these distributions. For example, the $(1-\alpha)$ confidence interval for a generic parameter $\theta$ is given by the empirical $\alpha/2$ and $(1-\alpha/2)$ quantiles of the corresponding bootstrap estimates $\{\widehat{\theta}^*_b\}_{b=1}^B$. This procedure provides a practical and robust approach to uncertainty quantification for all components of the model, including initial state probabilities, transition matrix entries, and state-specific precision parameters. All independent bootstrap computations can be performed in parallel to accelerate the inference process.

\subsection{Viterbi Algorithm}\label{sec:5.4}

The most probable sequence of hidden states $\widehat{z}_{1:T} = (\widehat{z}_1, \ldots, \widehat{z}_T)$ is obtained using the Viterbi algorithm \citep{Viterbi67}, which maximizes the joint posterior probability over paths:
\begin{equation*}
\widehat{z}_{1:T} = \arg\max_{z_{1:T}} \log P(z_{1:T} \mid y_{1:T}).
\end{equation*}

Define the recursion as
\begin{align*}
\delta_1(k) &= \log \pi_k + \log p(y_1 \mid z_1 = k),\\
\delta_t(j) &= \max_{i \in \{1, \ldots, K\}} \left[ \delta_{t-1}(i) + \log A_{ij} \right] + \log p(y_t \mid z_t = j), \quad t = 2, \ldots, T,\\
\psi_t(j) &= \arg\max_{i \in \{1, \ldots, K\}} \left[ \delta_{t-1}(i) + \log A_{ij} \right].
\end{align*}

The optimal path is then recovered by backtracking:
\begin{equation*}
\widehat{z}_T = \arg\max_{k \in \{1, \ldots, K\}} \delta_T(k), \quad
\widehat{z}_t = \psi_{t+1}(\widehat{z}_{t+1}), \quad t = T-1, \ldots, 1.
\end{equation*}

Table in Appendix~\ref{sec:Appendix_B} summarizes the pipeline for parameter estimation and model selection.

To ensure numerical stability, observed proportions and fitted means are clipped to $[\epsilon, 1-\epsilon]$ with $\epsilon = 10^{-8}$ before evaluating the Beta $\log$-density. 
The precision parameters are constrained to $[\phi_{\min}, \phi_{\max}]$ during optimization, with $\phi_{\min} = 1$ throughout and $\phi_{\max}$ set according to the regime of concentration expected in the data (default $\phi_{\max} = 500$ for the simulation design of Section~\ref{sec:6}, and $\phi_{\max} = 2{,}000$ for the mortality application of Section~\ref{sec:7}; 
see the diagnostic criterion introduced in Section~\ref{Sec:5}. The diagnostic thresholds $\Delta_{\text{abs}}$ and $\Delta_{\text{sum}}$ scale with $\phi_{\max}$ and are reported with the corresponding applications. Forward-backward recursions are carried out entirely on the $\log$ scale using the $\log$-sum-exp identity to prevent underflow. The likelihood of an HMM is invariant to permutations of the state labels \citep{ZML16}. To ensure identifiability across multi-start runs, the estimated states are reordered by increasing precision $\widehat\phi_1 < \cdots < \widehat\phi_K$ after each fit. For bootstrap and Monte Carlo replicates, estimated parameters are additionally aligned to the reference solution by minimizing the integrated squared  distance between the fitted and reference mean curves over all state 
permutations.


\section{Simulation Study}\label{sec:6}

To assess the finite-sample performance of the proposed Beta-GAM Hidden Markov Model, we conducted a Monte Carlo simulation study following the experimental design in \cite{LKGM17}.

The latent state sequence $\{z_t\}_{t=1}^T$ is generated from a homogeneous $K=4$-state Markov chain with transition matrix $\mathbf{A}_{\mathrm{true}}$ having diagonal entries equal to $0.95$ and off-diagonal entries equal to $0.05/3 \approx 0.0167$. The sample size is $T = 2{,}500$.

The covariate $x_t$ is drawn independently from $\mathrm{Unif}(-2,2)$. Conditionally on $z_t = k$, the response $y_t$ is generated from the $\mathrm{Beta}(\mu_{kt}\phi_k,\,(1-\mu_{kt})\phi_k)$ distribution, where the conditional mean $\mu_{kt} = \mathrm{logit}^{-1}(\eta_{kt})$ is obtained from a state-specific smooth function via a cubic B-spline basis with $M_{\mathrm{inner}}=6$ inner knots  \citep{EM21}. The four true mean functions include sinusoidal, quadratic, and exponential components, ensuring diverse functional forms across states. The precision parameters are $\bm{\phi}_{\mathrm{true}} = (10, 18, 28, 40)^{\top}$, representing increasing concentration from State~1 to State~4.


The simulation study is organized in two steps. First, we illustrate the complete model-selection pipeline on one representative simulated dataset. In this step, the number of states is searched over $K \in \{2,3,4,5\}$ and the smoothing parameter over
\[
\Lambda=\{0.005,0.01,0.05,0.1,0.5,1,5,10\},
\]
using $n_{\mathrm{start}}=15$ random initializations for each pair $(K,\lambda)$. The final model is selected by the two-stage procedure described in Section~\ref{Sec:5}, with $\mathrm{AIC}_{\nu}$ used to select $\lambda$ within each $K$, $\mathrm{BIC}_{\nu}$ used to select $K$, and diagnostic filtering used to remove degenerate solutions.

Second, following the simulation strategy of \cite{LKGM17}, we conduct a Monte Carlo experiment with $K$ fixed at its data-generating value, $K=4$. In each Monte Carlo replication, the smoothing parameter is selected over the same grid $\Lambda$ using $\mathrm{AIC}_{\nu}$, again with $n_{\mathrm{start}}=15$ random initializations for each candidate $\lambda$. This second step is intended to assess finite-sample estimation accuracy, smoothing-parameter selection, and latent-state decoding conditional on the correct number of states, rather than the repeated-sampling performance of the $K$-selection procedure.


To illustrate the model-selection pipeline in detail, we first present results from a representative single dataset. For this dataset, the two-stage selection procedure selects $K^{\ast}=4$ states and $\lambda^{\ast}=5$, matching the data-generating number of states. All 15 random initializations converge, indicating a well-behaved likelihood surface.

The estimated precision parameters are $\widehat{\bm{\phi}} = (9.40,\; 17.14,\; 32.42,\; 41.45)$, close to the true values $(10, 18, 28, 40)$, with a root mean squared error of $2.39$. The estimated transition matrix is strongly diagonal:

\[
\widehat{\mathbf{A}} =
\begin{pmatrix}
0.953 & 0.011 & 0.020 & 0.016 \\
0.014 & 0.953 & 0.019 & 0.013 \\
0.024 & 0.016 & 0.945 & 0.015 \\
0.012 & 0.024 & 0.004 & 0.959
\end{pmatrix},
\]
with diagonal entries ranging from 0.945 to 0.959, closely tracking the true value of 0.95 (RMSE $= 0.006$). The Viterbi decoding accuracy is 94.2\%, and the integrated curve RMSE is $0.008$.


\begin{table}[!htb]
\tabcolsep 0.4in
\centering
\caption{\small Model comparison across candidate $K$ values. For each $K$, 
the smoothing parameter $\lambda$ is selected by $\mathrm{AIC}_\nu$; 
criteria are reported for the best $\lambda$ within each $K$. 
All $K=5$ models were removed by the diagnostic 
filter.}\label{tab:selection_by_K}
\begin{tabular}{@{}ccrrr@{}}
\toprule
$K$ & $\lambda^*(K)$ & $\mathrm{AIC}_\nu$ & $\mathrm{BIC}_\nu$ & $\mathrm{ICL}_\nu$ \\
\midrule
2 & 1 & $-2810.1$ & $-2711.6$ & $-2364.9$ \\
3 & 5 & $-3415.9$ & $-3274.6$ & $-2719.5$ \\
\textbf{4} & \textbf{5} & $\mathbf{-3743.3}$ & $\mathbf{-3533.1}$ & $\mathbf{-2805.9}$ \\
5 & --- & \multicolumn{3}{c}{\textit{all removed by diagnostic filter}} \\
\bottomrule
\end{tabular}
\end{table}

Table~\ref{tab:selection_by_K} reports the information criteria for the best model within each candidate $K$, after diagnostic filtering and two-stage $\lambda$ selection, for this representative dataset. All $K=5$ models were discarded by the diagnostic filter due to insufficient state occupancy. Among the surviving candidates, $K=4$ achieves the lowest value of all three criteria by a wide margin. This table is intended to illustrate the proposed selection pipeline; the Monte Carlo assessment below is instead conducted conditional on the data-generating value $K=4$.

\paragraph{Monte Carlo results}

A total of $R = 100$ independent sequences of length $T=2,500$ datasets were generated and analyzed. Note that, as in \cite{LKGM17}, the Monte Carlo study is not designed to evaluate the repeated-sampling performance of the $K$-selection step. Rather, we use one representative simulated dataset to illustrate the full model-selection pipeline (see Table~\ref{tab:selection_by_K}), and then conduct the Monte Carlo experiment with $K$ fixed at the data-generating value. The objective of the Monte Carlo study is therefore to assess estimation accuracy, smoothing-parameter selection, and Viterbi decoding conditionally on the correct $K$, but not to evaluate the frequency with which the model-selection procedure recovers $K = 4$.

Table~\ref{tab:mc_summary} reports the main performance metrics, where the smoothing parameter $\lambda$ was selected as $\lambda = 5$ in 71\% of replications and $\lambda = 1$ in the remaining 29\%, confirming that the two-stage selection procedure identifies the appropriate level of smoothness and that the optimum lies well within the grid\footnote{For computational tractability, we adopt a common smoothing 
parameter $\lambda$ across all states, rather than state-specific 
parameters $\lambda_k$ as in \cite{LKGM17}. This choice reduces the dimensionality of the grid search from $|\Lambda|^K$ to $|\Lambda|$ and proved adequate in our simulation setting, where the effective degrees of freedom are similar across states. State-specific smoothing parameters are a natural extension that goes beyond the scope of this article. Note that in the Monte Carlo study we do not run a bootstrap inside each replication, but rather we use the simulations to assess estimation. Additional insights, such as the empirical coverage and average interval width, would require a nested Monte Carlo-bootstrap exercise, which is computationally infeasible.}

\begin{table}[!htb]
\centering
\tabcolsep 1.5in
\caption{\small Monte Carlo performance summary over 100 replications, conditional on the data-generating number of states ($K=4$, $T=2{,}500$, $\delta=0.95$). Standard deviations are reported in parentheses.}
\label{tab:mc_summary}
\begin{tabular}{@{}lr@{}}
\toprule
Metric                      & Mean (SD) \\
\midrule
Mean curve RMSE             & 0.011 (0.002) \\
Mean $\phi$ RMSE            & 1.75 (0.94) \\
Mean $A$ RMSE               & 0.008 (0.002) \\
Mean Viterbi accuracy (\%)  & 93.2 (1.1) \\
\bottomrule
\end{tabular}
\end{table}

Results were stable across $M_{\mathrm{inner}} = 4$ and $M_{\mathrm{inner}} = 6$ inner knots, consistent with the well-known insensitivity of P-spline fits to the number of basis elements once a sufficient threshold is reached \citep{EM96}.

\begin{figure}[!htb]
\centering
\includegraphics[width=0.85\linewidth]{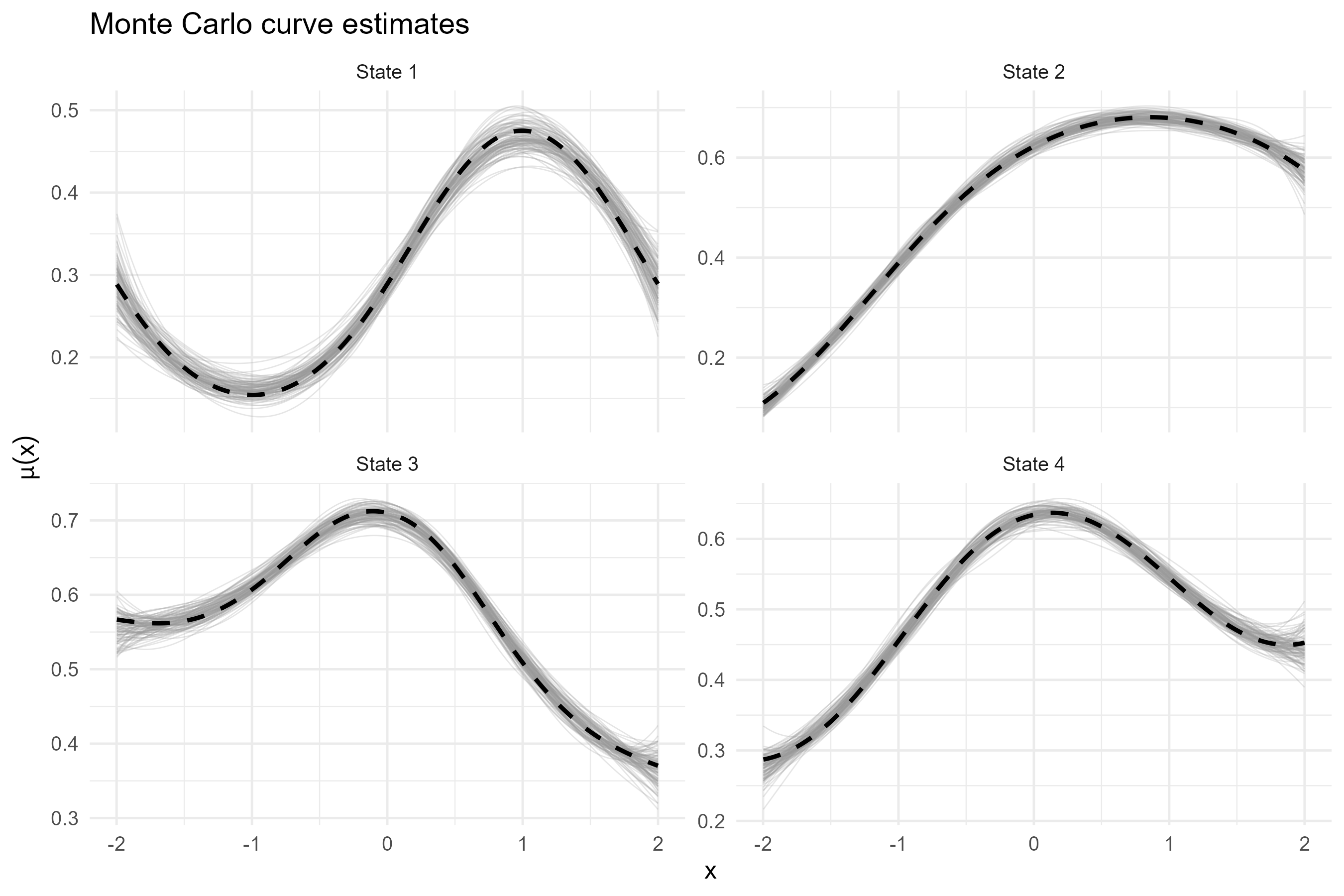}
\caption{\small Estimated state-specific mean curves from 100 Monte Carlo replications (thin grey lines), with the true functions shown as dashed black lines ($K=4$, $T=2{,}500$, $\delta=0.95$).}\label{fig:mc_curves}
\end{figure}

Figure~\ref{fig:mc_curves} displays the estimated state-specific mean curves from all 100 replications, overlaid on the true functions (dashed). The variability across replications is moderate and centered on the truth, confirming the consistency of the estimator. As expected, the dispersion is largest for State~1, which combines the lowest precision ($\phi_1 = 10$) with correspondingly noisier emissions, and increases at the boundaries of the covariate domain due to the reduced effective sample size near the edges of the support. The remaining three states exhibit progressively tighter concentration around the truth, consistent with their higher precision values.

To assess the sensitivity of the estimation procedure to reduced sample size and lower state persistence, we repeated the Monte Carlo study under a more challenging configuration with $T = 1{,}500$ and $\delta = 0.85$ (mean sojourn time $\approx 7$). 
Results are reported in Appendix~\ref{sec:Appendix_C}. As expected, all performance metrics worsen: the mean Viterbi accuracy decreases from 93.2\% to 80.7\%, the mean curve RMSE increases from 0.011 to 0.017, and the estimated mean curves do indeed show more variability (see Figure~\ref{fig:1500mc_curves} in Appendix~\ref{sec:Appendix_C}). Nevertheless, the estimation procedure remains stable, with 95 out of 100 replications converging successfully and the smoothing parameter selection shifting toward lower $\lambda$ values to accommodate the reduced effective sample size per state.


\section{Application to age-specific mortality ratios}\label{sec:7}

Sex differentials in mortality are a central theme in demography and public health. Across most populations, females exhibit lower mortality than males at nearly every age, but the magnitude of this advantage varies substantially with age and has evolved over time in response to epidemiological, behavioral, and social changes \citep[see, e.g.,][]{PLC19}. At young adult ages (roughly 15--35), the female advantage is particularly pronounced due to elevated male mortality from external causes such as accidents, violence, and alcohol-related deaths; at very young and older ages, the differential narrows. Detecting and characterizing shifts in these age-specific patterns is important for mortality forecasting, pension planning, and the design of age-targeted public health interventions.

Russia provides a particularly informative case study. The country experienced dramatic shifts in mortality over the second half of the twentieth century, driven by the late Soviet health crisis \citep{SMV96}, the transition shock of the early 1990s, the alcohol epidemic \citep{LCS97}, and the subsequent partial recovery. These structural changes offer a natural setting in which to test whether the proposed Beta-GAM HMM model can identify distinct mortality regimes from age-specific sex-ratio data.

The female-to-total mortality ratio $p_{x,t}$ provides a summary of the sex differential at each age $x$ and calendar year $t$. Because this ratio is bounded in $(0,1)$ and displays smooth but potentially nonlinear variation over age, it is well-suited to the proposed framework. Latent states can be interpreted as distinct mortality regimes, possibly associated with external factors. The restriction to ages 0--40 is motivated by the fact that the sex differential is most variable and most strongly shaped by behavioral and external causes at younger ages; at older ages, the ratio stabilizes and is dominated by chronic disease mortality, which is less amenable to regime-switching dynamics.

The data are from the Human Mortality Database (HMD). We extracted annual mortality counts for Russia spanning the period 1960--2014 (the full range available at the time of analysis) and focused on ages 0 to 40, yielding $T = 2{,}255$ age-year observations. The response variable is the proportion of female deaths within each age-year cell,
\[
p_{x,t} = \frac{D^{(f)}_{x,t}}{D_{x,t}},
\]
where $D^{(f)}_{x,t}$ denotes female deaths at age $x$ and year $t$, and $D_{x,t}$ the total.

\subsection{Model selection and estimation}

The estimation pipeline follows the procedure described in Section~\ref{Sec:5}. 

The number of states was searched over $K \in \{2, 3, 4, 5\}$ and the smoothing parameter over $\lambda \in \{0.1, 0.5, 1, 5, 10\}$, with $n_{\mathrm{start}} = 15$ random initializations per combination and a maximum of 300 EM iterations. Because Russian female-to-total mortality ratios are highly concentrated around their smooth age profile, we raised the precision bound to $\phi_{\max} = 2{,}000$ and rescaled the diagnostic thresholds, setting $\Delta_{\text{abs}} = 500$ and $\Delta_{\text{sum}} = 1{,}000$. Under these settings the estimated precisions $\widehat{\phi}_1 = 311.4$ and $\widehat{\phi}_2 = 643.9$ lie well below the bound, and the upper bootstrap endpoint (688.3 for $\phi_2$) is more than a factor of two below $\phi_{\max}$, confirming that the constraint is non-binding.

Table~\ref{tab:app_selection} reports the information criteria for the best model within each candidate $K$, after diagnostic filtering and two-stage $\lambda$ selection. Only the two-state model survives the joint convergence, non-degeneracy, and occupancy filters. Specifications with $K \geq 3$ are rejected because the additional states either fail to achieve sufficient posterior occupancy (below the 5\% threshold) or exhibit precision parameters that trigger the jump-detection filter, indicating over-splitting of the latent structure.

\begin{table}[!htb]
\centering
\caption{Model comparison across candidate $K$ values for the Russian mortality application. Only $K=2$ survives the diagnostic filter; models with $K \geq 3$ are rejected due to insufficient state occupancy or precision-based degeneracy.}\label{tab:app_selection}
\begin{tabular}{ccrrr}
\toprule
$K$ & $\lambda^*(K)$ & $\mathrm{AIC}_\nu$ & $\mathrm{BIC}_\nu$ & $\mathrm{ICL}_\nu$ \\
\midrule
\textbf{2} & \textbf{1} & $\mathbf{-11293.6}$ & $\mathbf{-11198.5}$ & $\mathbf{-10894.5}$ \\
3 & --- & \multicolumn{3}{c}{\textit{rejected by diagnostic filter}} \\
4 & --- & \multicolumn{3}{c}{\textit{rejected by diagnostic filter}} \\
5 & --- & \multicolumn{3}{c}{\textit{rejected by diagnostic filter}} \\
\bottomrule
\end{tabular}
\end{table}

The selected model has $K^* = 2$ states and $\lambda^* = 1$. All random initializations converge with a negligible gap between the best and median $\log$-likelihood, indicating a well-behaved likelihood surface.

\paragraph{Estimated parameters}

The estimated precision parameters are
\[
\widehat{\phi}_1 = 311.4, \qquad \widehat{\phi}_2 = 643.9,
\]
indicating two regimes with distinct levels of concentration: a more dispersed regime (State~1) and a highly concentrated regime (State~2). The estimated transition matrix is
\[
\widehat{\mathbf{A}} =
\begin{pmatrix}
0.962 & 0.038 \\
0.009 & 0.991
\end{pmatrix},
\]
with high diagonal entries indicating strong regime persistence in both states. Table~\ref{tab:app_ci} reports the 95\% parametric bootstrap confidence intervals based on $B = 200$ replicates (100\% convergence, no degenerate fits).
\begin{table}[!htb]
\centering
\tabcolsep 0.7in
\caption{Parameter estimates with 95\% parametric bootstrap confidence intervals ($B=200$) for the Russian mortality application.}\label{tab:app_ci}
\begin{tabular}{@{}lccc@{}}
\toprule
Parameter & Estimate & 95\% CI \\
\midrule
$\phi_1$ & 311.4 & $[280.4,\ 368.6]$ \\
$\phi_2$ & 643.9 & $[603.4,\ 688.3]$ \\
$A_{11}$ & 0.962 & $[0.942,\ 0.978]$ \\
$A_{12}$ & 0.038 & $[0.022,\ 0.058]$ \\
$A_{21}$ & 0.009 & $[0.005,\ 0.014]$ \\
$A_{22}$ & 0.991 & $[0.986,\ 0.995]$ \\
\bottomrule
\end{tabular}
\end{table}

The bootstrap intervals are narrow and well-separated across regimes: the two precision parameters have non-overlapping 95\% intervals ($[280.4,\,368.6]$ versus $[603.4,\,688.3]$), confirming that the two regimes differ in dispersion by a clearly identified factor. The transition probabilities are estimated with high precision --- the half-widths of the intervals for $A_{11}$ and $A_{22}$ are both below 0.02 --- and all intervals lie strictly inside $(0,1)$, with no boundary effects. Across the $B = 200$ bootstrap replicates, all fits converged and none was flagged by the diagnostic filter, so the reported quantiles are based on the full set of replicates.

\subsection{Interpretation of Regimes}

\begin{figure}[!htb]
\centering
\includegraphics[width=.9\linewidth]{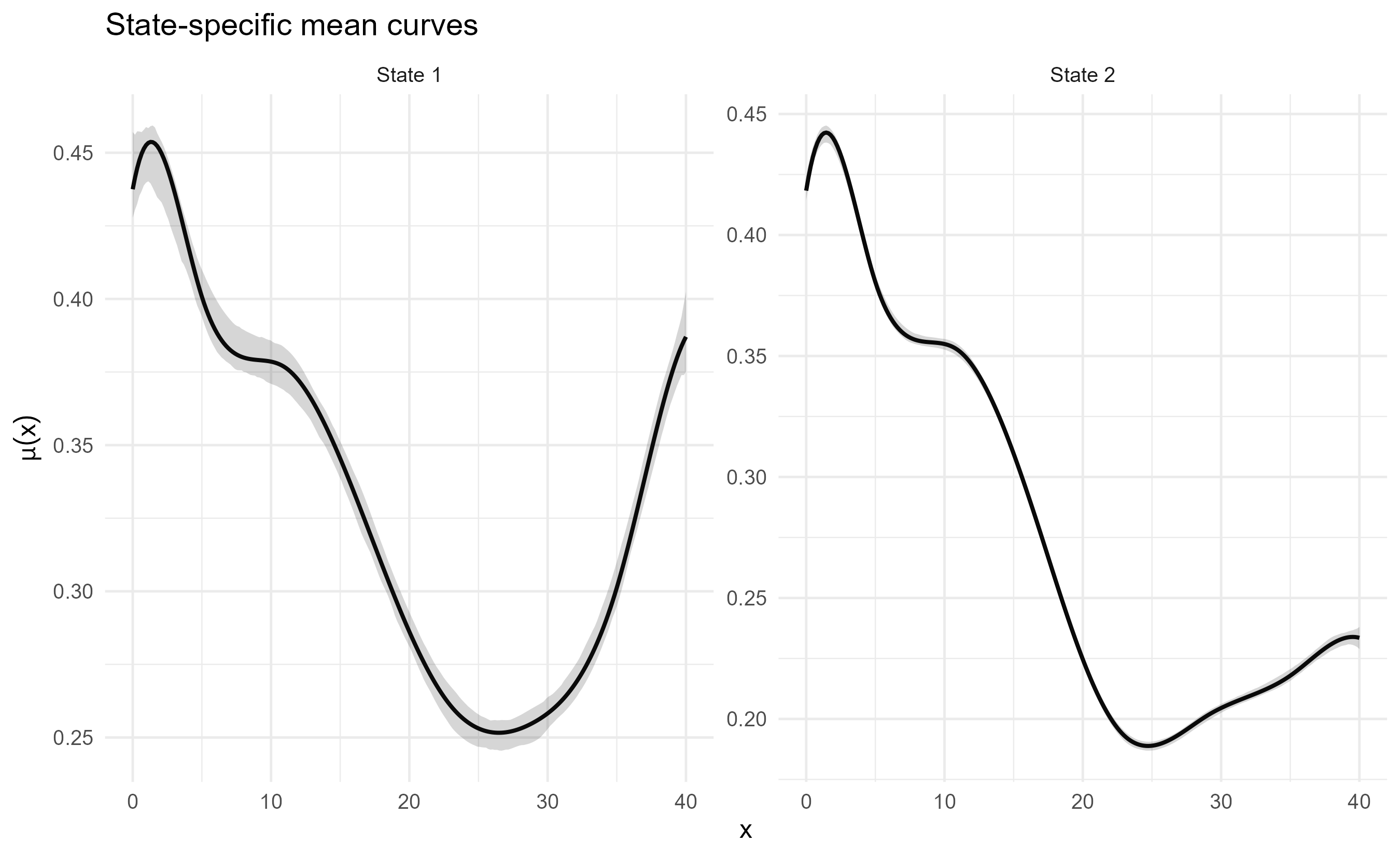}
\caption{Estimated state-specific age profiles of the female-to-total mortality ratio with 95\% pointwise bootstrap confidence bands (Russia, 1960--2014, ages 0--40, $B=200$).}\label{fig:app_curves}
\end{figure}

Figure~\ref{fig:app_curves} displays the estimated state-specific mean curves as a function of age, with 95\% pointwise bootstrap confidence bands. Both regimes share the characteristic profile of the female-to-total mortality ratio, with a pronounced depression in the 15--35 age range reflecting excess male mortality from external causes. The two regimes differ primarily in the depth of this depression and in the level of the ratio at working ages:

\begin{itemize}
\item \textbf{State~2} ($\widehat\phi_1 = 311$): a regime with a deeper trough in the young adult age range and higher dispersion, consistent with periods of elevated and more volatile male excess mortality.
\item \textbf{State~1} ($\widehat\phi_2 = 644$): a regime with a shallower trough and much higher precision, corresponding to a more stable and less male-skewed mortality pattern.
\end{itemize}

The Viterbi-decoded state sequence should be interpreted as a model-based summary of the latent regime structure, rather than as evidence of precisely dated structural breaks. The decoded path suggests a broad transition from the more dispersed regime to the more concentrated regime during the late 1960s, followed by a long period in which the more concentrated regime accounts for most age-year observations. Short-lived returns to the more dispersed regime are detected in the early 2010s. This pattern is broadly consistent with the historical evolution of Russian mortality, but the exact timing of the switches is conditional on the selected value of $K$, the smoothing parameter $\lambda$, and the use of the Viterbi decoding criterion.

\section{Conclusion}\label{sec:8}

We introduced a Hidden Markov Model with Beta emissions in which the mean is modeled by a GAM with $B$-spline basis. Estimation proceeds through a penalized EM algorithm, with bootstrap procedure to quantify uncertainty. The setup is well-suited to proportions, and it enables an interpretable separation between latent regimes driving the dynamics and smoothly varying covariate effects on the emission mean. Compared to the Bayesian Beta-HMM of \cite{CES22}, which uses state-specific but covariate-free emission parameters, our framework accommodates smooth nonlinear covariate effects within each regime and provides a diagnostic filter for detecting degenerate solutions that arise when the number of states is overspecified. The practical workflow consists of grid search over $(K,\lambda)$ combined with diagnostics to rule out degenerate solutions, and it provides a path to selecting stable and parsimonious models.

Simulation results suggest accurate recovery of the transition structure and state precisions: point estimates are close to the generating values, with full convergence of the bootstrap fits and no occurrence of degenerate solutions. In this controlled setting, where the true latent trajectory is known, Viterbi decoding recovers it with mean accuracy above 93\% under the baseline configuration, supporting the reliability of the decoded sequence as a summary of regime dynamics. 

In the application to the HMD data for Russia (1960--2014, ages 0--40), the selected two-state model yields a strongly diagonal transition matrix (high persistence in both regimes, with diagonal entries above 0.96) and state precisions that differ by roughly a factor of two ($\widehat{\phi}_1 = 311$, $\widehat{\phi}_2 = 644$), consistent with a concentrated distribution around the estimated age profiles. The two latent regimes admit a substantive demographic interpretation in terms of distinct age profiles of the female-to-total mortality ratio: a more dispersed regime with a deeper young-adult trough, dominant in the early 1960s and reappearing briefly in the early 2010s, and a more concentrated regime with a shallower trough, prevailing from the late 1960s onward. This decoded sequence aligns with the known history of Russian mortality, including the late Soviet health crisis, the transition shock of the early 1990s, and the alcohol epidemic. The fitted spline functions display smooth patterns over age, providing a compact summary of latent mortality structures by sex. Together, these features make the model a useful tool for describing and monitoring structural changes in a series of demographic proportions. 

The bootstrap intervals are narrow relative to the point estimates (relative half-width below 15\% for both state precisions) and strictly bounded away from the admissible boundaries for all transition probabilities, indicating a well-behaved uncertainty quantification on real data. The Viterbi path shows clear regime separation and highly persistent decoded regimes, with the chronological alternation reported in Section~\ref{sec:7} admitting a coherent demographic reading. Together with the simulations, these features suggest a well-calibrated pipeline with particular robustness in capturing regime persistence.

Several extensions are worth pursuing, and we briefly list three: 
\begin{inparaenum}[1)]
    \item Applying the framework to additional countries, age ranges, and time windows would help assess transferability and out-of-sample predictive performance. 
    \item A hierarchical formulation that shares information across populations (e.g., multiple countries) would be of considerable practical interest and is a natural direction for future work. 
    \item The model selection framework could be enriched by considering model averaging approaches as an alternative to the current pure selection strategy.   
\end{inparaenum} 
In general, the proposed Beta--GAM HMM strikes a practical balance among flexibility, interpretability, and computational tractability to analyze proportion data.

\section*{Acknowledgment}

The second author acknowledges financial support from an Australian Research Council Future Fellowship (ARC FT240100338), and thanks insightful comments and suggestions from seminar participants of the Dondena seminar series at Bocconi University in 2025. The third author was supported in part by the MUSA -- Multilayered Urban Sustainability Action project, funded by the European Union -- Next Generation EU, under the National Recovery and Resilience Plan (NRRP) Mission 4 Component 2 Investment line 1.5.

\section*{Data availability statement}
The data that support the findings of this study are openly available in \url{https://www.mortality.org/}.

\vspace{-.15in}

\section*{Disclosure Statement}
We declare that the authors have no relevant financial or non-financial competing interests to report.

\newpage
\appendix

\section{Derivatives of the Penalized Log-Likelihood for the Beta-GAM-HMM Model}\label{sec:Appendix_A}

Consider the emission model:
\[
y_t \mid z_t = k \sim \mathrm{Beta}(a_{kt}, b_{kt}),
\]
with the reparameterization:
\[
a_{kt} = \mu_{kt} \phi_k, \quad b_{kt} = (1 - \mu_{kt}) \phi_k,
\]
where \( \mu_{kt} \in (0, 1) \) is the mean, modeled via a smooth additive predictor, and \( \phi_k > 0 \) is the precision parameter for state \( k \).

The $\log$-density of a single observation \( y_t \in (0,1) \), conditional on \( z_t = k \), is:
\[
\begin{aligned}
\log p(y_t \mid z_t = k) =\;&
\log \Gamma(\phi_k) 
- \log \Gamma(\mu_{kt} \phi_k)
- \log \Gamma((1 - \mu_{kt}) \phi_k) \\
&+ (\mu_{kt} \phi_k - 1) \log y_t
+ ((1 - \mu_{kt}) \phi_k - 1) \log(1 - y_t),
\end{aligned}
\]
where \( \Gamma(\cdot) \) denotes the Gamma function.

The mean is linked to the linear predictor \( \eta_{kt} \) via:
\[
\mu_{kt} = \frac{1}{1 + \exp(-\eta_{kt})}, \quad \text{with} \quad \eta_{kt} = B_t^\top \beta_k,
\]
where \(B_t \in \mathbb{R}^{pM} \) is the spline basis vector evaluated at time \( t \), and \(\beta_k \in \mathbb{R}^{pM} \) is the coefficient vector for state \( k \).

\paragraph{Gradient with Respect to $\beta_k$}

Let
\[
\dot{\mu}_{kt}
=
\frac{\partial \mu_{kt}}{\partial \eta_{kt}}
=
\mu_{kt}(1-\mu_{kt}).
\]
Since $\eta_{kt}=\mathbf{B}_t^\top\beta_k$, we also have
\[
\frac{\partial \eta_{kt}}{\partial \beta_k} = \mathbf{B}_t.
\]
Therefore, by the chain rule,
\[
\frac{\partial \mu_{kt}}{\partial \beta_k}
=
\dot{\mu}_{kt}\,\mathbf{B}_t.
\]

The gradient of the (unpenalized) $\log$-likelihood with respect to \( \beta_k \) is:
\[
\nabla_{\beta_k} \log p(y_t \mid z_t = k)
=
\phi_k
\left[
\log\!\left(\frac{y_t}{1-y_t}\right)
-\psi(\mu_{kt}\phi_k)
+\psi((1-\mu_{kt})\phi_k)
\right]
\dot{\mu}_{kt}\,\mathbf{B}_t,
\]
where \( \psi(\cdot) \) denotes the digamma function.

When the P-spline difference penalty $\lambda\,\beta_k^\top \mathbf{P}\,\beta_k$, with $\mathbf{P} = \mathbf{D}_d^\top \mathbf{D}_d$, is added, the penalized expected $\log$-likelihood becomes:
\[
\widetilde{\ell}_k(\beta_k) = \sum_{t=1}^T \gamma_{tk} \log p(y_t \mid z_t = k) - \lambda \beta_k^\top \mathbf{P}\,\beta_k.
\]

The corresponding penalized gradient is:
\[
\nabla_{\beta_k} \widetilde{\ell}_k = \sum_{t=1}^T \gamma_{tk} \nabla_{\beta_k} \log p(y_t \mid z_t = k) - 2\lambda \mathbf{P} \beta_k.
\]

\paragraph{Gradient with Respect to $\phi_k$}

The gradient of the $\log$-likelihood with respect to the precision parameter \( \phi_k \) is:
\begin{equation*}
\begin{aligned}
\frac{\partial}{\partial \phi_k} \log p(y_t \mid z_t = k) =\;&
\psi(\phi_k)
- \mu_{kt} \psi(\mu_{kt} \phi_k)
- (1 - \mu_{kt}) \psi((1 - \mu_{kt}) \phi_k) \\
&+ \mu_{kt} \log y_t
+ (1 - \mu_{kt}) \log(1 - y_t).
\end{aligned}
\end{equation*}

This derivative is used in the M-step to update $\phi_k$ via numerical optimization. Note that the penalization does not affect the update of $\phi_k$, as it depends only on $\beta_k$.

\section{Estimation and model selection for the Beta-GAM HMM}\label{sec:Appendix_B}

\begin{center}
\small
\noindent
\fbox{%
\parbox{\linewidth}{%

\begin{enumerate}
\item \textbf{Grid definition.}
Specify the candidate numbers of latent states
$\mathcal{K}=\{2,\ldots,K_{\max}\}$ and the smoothing penalty grid
$\Lambda=\{\lambda_1,\ldots,\lambda_L\}$, and define $\mathcal{G}=\mathcal{K}\times\Lambda$.
\item \textbf{Penalized EM fitting.}
For each $(K,\lambda)\in\mathcal{G}$, fit the Beta-GAM HMM using multiple
random initializations and retain the solution with the largest observed
log-likelihood. At each EM iteration:

\begin{itemize}
\item \textit{E-step.} For each state $k$, compute $\eta_{kt}=\mathcal{B}_t^\top\beta_k, \mu_{kt}=\operatorname{logit}^{-1}(\eta_{kt})$,
and evaluate the Beta emission density $y_t\mid z_t=k
    \sim
    \operatorname{Beta}\bigl(\mu_{kt}\phi_k,\,(1-\mu_{kt})\phi_k\bigr)$.
The posterior probabilities $\gamma_{tk}$ and $\xi_{tij}$ are then obtained
by the forward--backward recursions.

\item \textit{M-step.} Update the initial and transition probabilities by
\[
    \widehat{\pi}_k=\gamma_{1k},
    \qquad
    \widehat{A}_{ij}
    =
    \frac{\sum_{t=1}^{T-1}\xi_{tij}}
         {\sum_{t=1}^{T-1}\gamma_{ti}} .
\]
For each state $k$, update $(\beta_k,\phi_k)$ by numerical maximization of the
penalized emission objective
\[
    Q_k(\beta_k,\phi_k)
    =
    \sum_{t=1}^T
    \gamma_{tk}\log p(y_t\mid \beta_k,\phi_k)
    -
    \lambda \beta_k^\top P\beta_k .
\]
\end{itemize}

\item \textbf{Information criteria.}
For each fitted model, compute the effective degrees of freedom $\nu$ from
the penalized hat-matrix approximation and evaluate
\[
    \mathrm{AIC}_\nu
    =
    -2\ell(\widehat\theta)+2\nu,
    \qquad
    \mathrm{BIC}_\nu
    =
    -2\ell(\widehat\theta)+\nu\log T,
\qquad
    \mathrm{ICL}_\nu
    =
    \mathrm{BIC}_\nu
    -
    2\sum_{t=1}^T\sum_{k=1}^K
    \gamma_{tk}\log\gamma_{tk}.
\]

\item \textbf{Diagnostic filtering.}
For each fitted model, order the precision estimates as
$\phi_{(1)}\leq\cdots\leq\phi_{(K)}$ and compute
\[
    N_{\mathrm{sat}}
    =
    \sum_{k=1}^K
    \mathbb{I}(\phi_k\geq \phi_{\max}-\epsilon),
    \qquad
    \Delta_i=\phi_{(i+1)}-\phi_{(i)}, \quad i=1,\ldots,K-1.
\]
Let $m=\min(3,K-1)$, $\Delta_{\mathrm{tail}} = \sum_{i=K-m}^{K-1}\Delta_i$. Discard the model as degenerate if $N_{\mathrm{sat}}\geq s_{\mathrm{thresh}}$, $\max_{i\in\{K-m,\ldots,K-1\}}\Delta_i>\Delta_{\mathrm{abs}}$,
    \text{or}\
    $\Delta_{\mathrm{tail}}>\Delta_{\mathrm{sum}}$. Equivalently, retain only models with diagnostic indicator $\mathcal{E}(K)=0$. The surviving set is denoted by $\mathcal{G}_{\mathrm{valid}} = \{(K,\lambda)\in\mathcal{G}:\mathcal{E}(K)=0\}$.

\item \textbf{Two-stage final selection.}
First, for each candidate $K$, select the smoothing parameter by
\[
    \lambda^*(K)
    =
    \arg\min_{\lambda:(K,\lambda)\in\mathcal{G}_{\mathrm{valid}}}
    \mathrm{AIC}_\nu(K,\lambda).
\]
Second, compute $\mathrm{BIC}_\nu\bigl(K,\lambda^*(K)\bigr)$ for each
surviving $K$, define $\mathrm{BIC}_{\nu,\min} = \min_K \mathrm{BIC}_\nu\bigl(K,\lambda^*(K)\bigr)$, and select
$K^* = \max \left\{K: \mathrm{BIC}_\nu\bigl(K,\lambda^*(K)\bigr) \leq \mathrm{BIC}_{\nu,\min}+2\right\}$. Return the final model $\bigl(K^*,\lambda^*(K^*)\bigr)$.
\end{enumerate}
}
}
\end{center}

\section{Robustness analysis}\label{sec:Appendix_C}

\begin{figure}[!htb]
\centering
\includegraphics[width=0.87\linewidth]{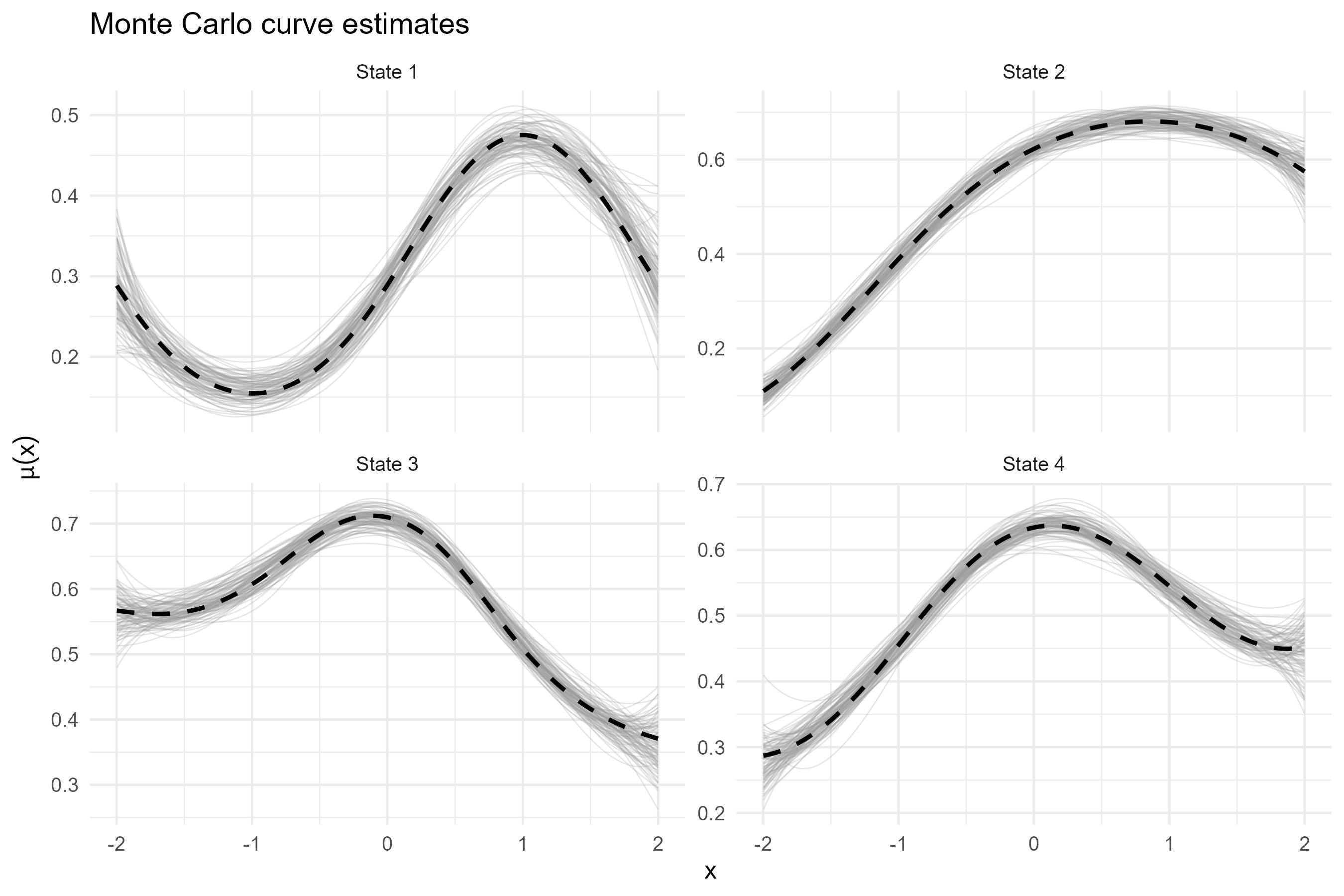}
\caption{\small Estimated state-specific mean curves from 100 Monte Carlo replications (thin grey lines), with the true functions shown as dashed black lines ($K=4$, $T=1{,}500$, $\delta=0.85$).}\label{fig:1500mc_curves}
\end{figure}

Table~\ref{tab:mc_robust} compares Monte Carlo performance under 
the baseline scenario ($T=2{,}500$, $\delta=0.95$) and a more 
challenging configuration ($T=1{,}500$, $\delta=0.85$, Figure~\ref{fig:1500mc_curves}, provides a graphical representation of the MC scenarios).
\begin{table}[!htb]
\centering
\caption{\small Monte Carlo performance comparison across simulation 
scenarios. Standard deviations in parentheses.}\label{tab:mc_robust}
\tabcolsep 0.6in
\renewcommand{\arraystretch}{0.8} 
\begin{tabular}{@{}lcc@{}}
\toprule
Metric & Scenario~1 & Scenario~2 \\
 & ($T=2500$, $\delta=0.95$) & ($T=1500$, $\delta=0.85$) \\
\midrule
Replications (valid/total) & 100/100 & 95/100 \\
Mean curve RMSE & 0.011 (0.002) & 0.017 (0.003) \\
Mean $\phi$ RMSE & 1.75 (0.94) & 3.51 (2.52) \\
Mean $A$ RMSE & 0.008 (0.002) & 0.025 (0.007) \\
Mean Viterbi accuracy (\%) & 93.2 (1.1) & 80.7 (2.2) \\
\midrule
$\lambda$ selected: $\lambda=5$ & 71\% & 29\% \\
$\lambda$ selected: $\lambda=1$ & 29\% & 56\% \\
$\lambda$ selected: other & 0\% & 15\% \\
\bottomrule
\end{tabular}
\end{table}

Under the challenging scenario, performance degrades in a predictable and interpretable manner. The lower state persistence ($\delta = 0.85$, mean sojourn $\approx 7$ versus $\approx 20$) reduces the effective number of contiguous observations per state visit, making it harder for the forward-backward algorithm to 
separate the states and for the spline estimates to borrow strength within regimes. The Viterbi accuracy drops by approximately 12 percentage points, and the precision parameters are estimated with roughly double the RMSE. The smoothing parameter selection adapts accordingly, shifting toward lower $\lambda$ values (56\% select 
$\lambda=1$ versus 71\% selecting $\lambda=5$ in the baseline), reflecting the need for more flexible curves when less information is available per state. Despite these challenges, the procedure remains operational, with 95\% of replications producing valid estimates.

\newpage
\bibliographystyle{agsm}
\bibliography{beta_gam_hmm}

@article{FC04,
    author = {S. Ferrari and F. {Cribari-Neto}},
    title = {Beta regression for modelling rates and proportions},
    journal = {Journal of Applied Statistics},
    year = {2004},
    volume = {31},
    number = {7},
    pages = {799-815}
}

@article{DW19,
    author = {J. C. Douma and J. T. Weedon},
    title = {Analysing continuous proportions in ecology and evolution: {A} practical introduction to beta and {D}irichlet regression},
    journal = {Methods in Ecology and Evolution},
    year = {2019},
    volume = {10},
    pages = {1412-1430}
}

@book{NW06,
    author = {J. Nocedal and S. J. Wright},
    title = {Numerical {O}ptimization},
    publisher = {Springer},
    year = {2006},
    edn = {2nd},
    address = {New York}
}

@article{BKC20,
    author = {U. Basellini and S. Kjaergaard and C. G. Camarda},
    title = {An age-at-death distribution approach to forecast cohort mortality},
    journal = {Insurance: Mathematics and Economics},
    year = {2020},
    volume = {91},
    pages = {129-143}
}

@article{PLC19,
    author = {M. D. Pascariu and A. Lenart and Y. {Canudas-Romo}},
    title = {The maximum entropy mortality model: Forecasting mortality using statistical moments},
    journal = {Scandinavian Actuarial Journal},
    year = {2019},
    volume = {2019},
    number = {8},
    pages = {661-685}
}

@article{KU01,
    author = {A. Kneip and K. Utikal},
    title = {Inference for density families using functional principal componenent analysis},
    journal = {Journal of the American Statistical Association: Theory and Methods},
    year = {2001},
    volume = {96},
    number = {454},
    pages = {519-542}
}

@article{KMP+19,
    author = {P. Kokoszka and H. Miao and A. Petersen and H. L. Shang},
    title = {Forecasting of density functions with an application to cross-sectional and intraday returns},
    journal = {International Journal of Forecasting},
    year = {2019},
    volume = {35},
    pages = {1304-1317}
}

@incollection{JCP+08,
    author = {W. Jank and G. Shmeuli and C. Plaisant and B. Shneiderman},
    title = {Visualizing functional data with an application to ebay's online auctions},
    booktitle = {Handbook of Data Visualization},
    publisher = {Springer-Verlag},
    year = {2008},
    address = {Berlin},
    editor = {C. Chen and W. H\"{a}rdle and A. Unwin}
}

@book{Wood17,
    author = {S. N. Wood},
    title = {Generalized Additive Models: {A}n Introduction with {R}},
    publisher = {Chapman and Hall/CRC},
    year = {2017},
    address = {New York}
}

@article{HYS+19,
    author = {F. K. C. Hui and C. You and H. L. Shang and S. M\"{u}ller},
    title = {Semiparametric regression using variational approximations},
    journal = {Journal of the American Statistical Association: Theory and Methods},
    year = {2019},
    volume = {14},
    number = {528},
    pages = {1765-1777}
}

@article{CES22,
  author  = {Can, Ceren Eda and Ergun, Gul and Soyer, Refik},
  title   = {Bayesian Analysis of Proportions via a Hidden {M}arkov Model},
  journal = {Methodology and Computing in Applied Probability},
  year    = {2022},
  volume  = {24},
  pages   = {3121--3139},
  doi     = {10.1007/s11009-022-09971-0}
}

@article{LKGM17,
  author  = {Langrock, Roland and Kneib, Thomas and Glennie, Richard and Michelot, Th\'{e}o},
  title   = {Markov-switching generalized additive models},
  journal = {Statistics and Computing},
  year    = {2017},
  volume  = {27},
  pages   = {259--270},
  doi     = {10.1007/s11222-015-9620-3}
}

@article{EM96,
   author  = {Eilers, Paul H. C. and Marx, Brian D.},
   title   = {Flexible smoothing with {B}-splines and penalties},
   journal = {Statistical Science},
   year    = {1996},
   volume  = {11},
 pages   = {89--121}   
 }

@book{EM21,
   author    = {Eilers, Paul H. C. and Marx, Brian D.},
   title     = {Practical Smoothing: The Joys of {P}-splines},
   publisher = {Cambridge University Press},
   year      = {2021},
   address = {Cambridge}
 }

@article{Gray92,
   author  = {Gray, Robert J.},
   title   = {Flexible methods for analyzing survival data using splines, with application to breast cancer prognosis},
   journal = {Journal of the American Statistical Association: {A}pplications and Case Studies},
   year    = {1992},
   volume  = {87},
   pages   = {942--951}
 }

@article{DLR77,
   author = {Dempster, A. P. and Laird, N. M. and Rubin, D. B.},
   title = {Maximum likelihood from incomplete data via the {EM} algorithm},
   journal = {Journal of the Royal Statistical Society: Series B},
   year = {1977},
   volume = {39},
   pages = {1--38}
 }

@article{SMV96,
  author  = {Shkolnikov, Vladimir and Mesl{\'e}, France and Vallin, Jacques},
  title   = {Health crisis in {R}ussia {I}. {R}ecent trends in life expectancy and causes of death from 1970 to 1993},
  journal = {Population: An English Selection},
  year    = {1996},
  volume  = {8},
  pages   = {123--154}
}

@book{ZML16,
    author = {W. Zucchini and I. L. MacDonald and R. Langrock},
    title = {Hidden Markov Models for Time Series: An Introduction Using R},
    publisher = {Chapman and Hall/CRC},
    year = {2016},
    edition = {2nd}
}

@article{Schwarz78,
    author = {Schwarz, G.},
    title = {Estimating the dimension of a model},
    journal = {Annals of Statistics},
    year = {1978},
    volume = {6},
    number = {2},
    pages = {461-464}
}

@article{Viterbi67,
    author = {Viterbi, A. J.},
    title = {Error bounds for convolutional codes and an asymptotically optimum decoding algorithm},
    journal = {IEEE Transactions on Information Theory},
    year = {1967},
    volume = {13},
    number = {2},
    pages = {260-269}
}

@article{BCG00,
    author = {Biernacki, C. and Celeux, G. and Govaert, G.},
    title = {Assessing a mixture model for clustering with the integrated completed likelihood},
    journal = {IEEE Transactions on Pattern Analysis and Machine Intelligence},
    year = {2000},
    volume = {22},
    number = {7},
    pages = {719-725}
}

@article{Akaike74,
    author = {H. Akaike},
    title = {A new look at the statistical model identification},
    journal = {IEEE Transactions on Automatic Control},
    year = {1974},
    volume = {19},
    number = {6},
    pages = {716-723}
}

@article{LCS97,
    author  = {Leon, David A. and Chenet, Laurent and Shkolnikov, Vladimir M. and Zakharov, Sergei and Shapiro, Judith and Rakhmanova, Galina and Vassin, Sergei and McKee, Martin},
    title   = {Huge variation in {R}ussian mortality rates 1984--94: {A}rtefact, alcohol, or what?},
    journal = {The Lancet},
    year    = {1997},
    volume  = {350},
    number  = {9075},
    pages   = {383--388}  
}

\end{document}